\documentclass[12pt]{article}
\begin{document}
\def\fin{\end{document}}
\newcommand{\be} {\begin{equation}}
\newcommand{\ee} {\end{equation}}
\newcommand{\ba} {\begin{eqnarray}}
\newcommand{\ea} {\end{eqnarray}}
\def\R{{\hbox{{\rm I}\kern-.2em\hbox{\rm R}}}}
\begin{titlepage}
\begin{flushright} UMH-MG-98/01\\ ULB-TH-98/06 
\end{flushright}
\vskip 1.cm
\begin{center} {\LARGE\bf G\"odel metric as a squashed anti--de Sitter geometry}\\
\vskip 1.cm  M. Rooman\footnote{Ma\^\i tre de Recherches F.N.R.S.}\strut '\footnote{E-mail
mrooman@ulb.ac.be}, \\
\vskip 0.5 cm
 {\em Service de Physique th\'eorique,}\\
 {\em Universit\'e Libre de Bruxelles}, \\ {\em  C.P. 225, bvd du Triomphe, B-1050 Bruxelles,
Belgium}
\vskip 0.75 cm
 Ph. Spindel\footnote{E-mail spindel@sun1.umh.ac.be}
\\
\vskip 0.5 cm
 {\em M\'ecanique et Gravitation}\\
 {\em Universit\'e de Mons-Hainaut}, \\ {\em 6, avenue du Champ de Mars, B-7000 Mons, Belgium}
\vskip 1 cm
\end{center}
\begin{abstract} We show that the non flat factor of the G\"odel metric
belongs to a one parameter family of 2+1 dimensional geometries that also includes the
anti--de Sitter metric. The elements of this family allow a generalization \`a la
Kaluza--Klein of the usual 3+1 dimensional G\"odel metric. Their lightcones can be viewed as
deformations of the anti--de Sitter ones, involving tilting and squashing. This provides a
simple geometric picture of the causal structure of these space--times, anti--de Sitter
geometry appearing as the boundary between causally safe and causally pathological spaces.
Furthermore, we construct a global algebraic isometric embedding of these metrics in  4+3 or
3+4 dimensional flat spaces, thereby illustrating in another way the occurrence of the closed
timelike curves.
\end{abstract}
\vfill 
\end{titlepage}

G\"odel's cosmological model \cite{God1,God2} has been among the most intriguing exact
solutions to Einstein equations \cite{KSHM} and remains still today quite interesting
both mathematically and physically. The feature that has
contributed most to its fame is the occurrence of closed timelike curves through each of its
points \cite{HwEl}. This  a priori limits its physical relevance, as
such spaces are usually unstable with respect to quantum fluctuations \cite{Hw}. However,
occurrences of causality breakdowns are intrinsically interesting to study (see for instance
the huge number of papers devoted to time machines \cite{Na}). Moreover, causal
pathologies are large scale deficiencies, and thus parts of these spaces surrounded by more
standard space--times can represent (rotating) objects with physical meaning
\cite{BSMcC,LRS}. 
 
In this paper, we focus on geometric properties of a one parameter family of 2+1
dimensional metrics connecting the non trivial part of the G\"odel metric to the anti--de
Sitter (AdS) geometry. More precisely, by expressing the metrics of this family in an
appropriate coframe, we show that they can be interpreted as resulting from a
directional squashing of the lightcones of AdS space. We give the explicit form of their
Killing vector fields, whose algebra corresponds to the breaking of the $so(2,2)$ AdS
isometry group into
$so(2,1)\times so(2)$. We then build a normal geodesic coordinate system so as to clarify the
geometric meaning of the space--times we consider. We finally obtain a picture of these spaces
as the intersection of 4 quadratic surfaces in a flat 7--dimensional space, thereby clarifying
the origin of the closed timelike curves. Note that the family of metrics we examine has been
considered from a different point of vue a long time ago as
homogeneous cylindrically metrics,  solutions of Einstein--Maxwell \cite{RTh}/
Einstein--Maxwell--scalar \cite{RTi} field equations, and more recently in the framework of a
low energy string effective action \cite{BD}.

The 3+1 dimensional G\"odel metric can be expressed
as the direct riemannian sum $dz^2+d\sigma^2$ of a flat factor and the 2+1 dimensional
metric~: 
\be d\sigma^2=-dt^2+dx^2-2\; e^{x/a}dtdy-\frac12 e^{2x/a}dy^2 \qquad .\label{dsG}
\ee This four dimensional geometry is $G_5$ invariant, and solves Einstein equations with as
sources a negative cosmological constant $\Lambda$ and a pressureless perfect fluid of
constant energy density (dust)~:  
\be G_{\mu\nu}+\Lambda g_{\mu\nu}=T_{\mu\nu}\qquad,\qquad T_{\mu\nu}=\rho u_\mu u_\nu \qquad
, \qquad \rho= -2\Lambda \qquad .
\ee Hereafter we shall restrict ourselves to the (non-trivial) three dimensional part of the
metric, $d\sigma^2$ [eq.(\ref{dsG})], which we shall still call the G\"odel metric. We embed
it in a one parameter family of geometries. To this end, let us introduce the triad~: 
\be
\theta^0=dt+e^{x/a}dy\quad,\quad \theta^1=dx\quad,\quad\theta^2=e^{x/a}dy\quad,\label{triad}
\ee and consider the set of metrics
\ba d\sigma^2_{\mu}&=&-({\bf\theta}^0)^2+(\theta^1)^2+\frac{1}{\mu^2}(\theta^2)^2 \nonumber\\
&=& -dt^2+dx^2-2\;e^{x/a}dtdy+\frac{1-\mu^2}{\mu^2} e^{2x/a}dy^2 \qquad .
\label{dsm}
\ea 
The G\"odel metric corresponds to $\mu^2=2$. The Ricci tensors of these metrics have as
non vanishing  components with respect to the basis (\ref{triad})~: 
\be R^0_0=-\frac{\mu^2}{2a^2}\quad,\quad R^1_1=R^2_2=\frac{\mu^2-2}{2a^2}\qquad.\label{Ricci}
\ee Hence, only the space with $\mu^2=1$ has constant (negative) curvature. It corresponds
(at least locally) to an AdS space. Note also that the Cotton-York tensors
of these spaces are diagonal~: 
\be C^0_0=\frac{\mu(1-\mu^2)}{a^3}\quad,\quad C^1_1=C^2_2=-\frac{\mu(1-\mu^2)}{2\;a^3}\quad ,
\ee and vanish only for $\mu^2=1$, indicating that, in this family, only the AdS metric is
conformally flat.  

A standard calculation shows that the metrics $d\sigma_{\mu}^2$ admit at least the four
Killing vector fields~:\\ 
\be
\begin{array}{l c l c}
 \xi_1=\partial_y \nonumber   &,&
 \xi_2=\partial_x-\frac ya\partial_y  \nonumber  &, \\
 \xi_3=-{\mu^2} e^{-x/a}\partial_t+\frac ya
\partial_x+(\frac{\mu^2}2e^{-2x/a}- \frac{y^2}{2\;a^2})\partial_y  \nonumber   &,&
 \xi_4=\partial_t  &.
\end{array}
\ee
\strut\\ These vectors obey the $so(2,1)\times so(2)$ algebra~: 
\ba [\xi_1,\xi_2]=-\frac 1a \xi_1 \quad ,&\quad [\xi_1,\xi_3]=+\frac 1a \xi_2 \quad ,&\quad 
[\xi_2,\xi_3]=-\frac 1a \xi_3 \quad ,\quad \nonumber\\
\strut [\xi_1,\xi_4] =0 
\quad ,&\quad [\xi_2,\xi_4]=0\quad ,&\quad  [\xi_3,\xi_4]=0 \quad ,
\ea for which a more standard representation is obtained by using as basis the vectors
$\xi_4$ and 
\be B_1^u=\frac a{\sqrt{2}}(\xi_1+\xi_3)\quad,\quad  B_1^v=\frac
a{\sqrt{2}}(\xi_1-\xi_3)\quad,\quad  L_1^z=a\xi_2 \quad .
\ee For the AdS space, two extra Killing vectors are found~: 
\ba &\xi_5=\sin (t/a)\partial_t-\cos (t/a) \partial_x-e^{-x/a} \sin (t/a)
\partial_y&\quad ,
\nonumber\\ &\xi_6=\cos (t/a)\partial_t+\sin (t/a) \partial_x-e^{-x/a} \cos (t/a) \partial_y&
\quad .
\ea These vectors commute with $\xi_1,\xi_2,\xi_3$ and extend the $so(2)$ subalgebra generated
by
$\xi_4$ to the second $so(2,1)$ factor of $so(2,2)$, the well known algebra of isometries of
AdS space~: 
\ba &[\xi_4,\xi_5]=+\frac 1a \xi_6 \quad,\quad [\xi_4,\xi_6]=-\frac 1a \xi_5 \quad,\quad 
[\xi_5,\xi_6]=-\frac 1a \xi_4 \quad .&
\ea

The metrics (\ref{dsm}) can be added to $(D-3)$--dimensional Euclidean Einstein metrics 
$d\Sigma_{\nu}^2$ (metrics of constant Ricci curvature~: $R^i_j=\nu \delta^i_j$)  to give a
$D$--dimensional generalization of the classical G\"odel solution~: 
$ds^2=d\sigma_{\mu}^2+d\Sigma_{\nu}^2$ with  the values of their parameters given by~:
\be a^2=\frac{D-2}{(D-4)\rho - 4\Lambda }\quad,\quad
\mu^2=2\frac{(D-3)\rho - 2\Lambda }{(D-4)\rho - 4\Lambda }\quad,\quad \nu=\frac{\rho +
2\Lambda }{D-2}\quad . \ee When $D>4$, the $(D-3)$ space dimensions added to the G\"odel
geometry are all of the same order of magnitude, and thus do not correspond to usual
Kaluza-Klein cosmologies, which require 4 non compact dimensions and $D-4$ compact ones. It
is nevertheless not excluded that this kind of solutions contain some elements of physical
relevance, for instance in the framework of the hot phase of the early universe, where
dimensional phase transitions could occur. When $D=4$, we have $\nu =0 $, and the flat factor
can be supposed compact (a circle) or not (a line). In the latter case, we reobtain the
original 4--dimensional G\"odel solution.  Note that, if we accept the positivity of the
energy density
$\rho$, the value of the cosmological constant is bounded from above~:  
\be
\Lambda<\frac{D-4}4 \rho \qquad ,
\ee in order to ensure the positivity of $a^2$ and $\mu^2$. Moreover, if we impose 
$\nu$ to be positive\footnote{Though compact spaces can have negative curvature, they are
obtained by identification of a non compact universal covering and thus look less natural.},
the values of $\Lambda$ are restricted to the interval $[-\rho /2,(D-4)\rho /4]$.

As shown above,  the metric (\ref{dsm}) becomes for $\mu^2 =1$~: 
\be d\sigma^2=-dt^2+dx^2-2 e^{x/a}dtdy  \label{dsAdS}
\ee and describes the geometry of an AdS space, in unusual coordinates. Our purpose  now is
to make the connection between these coordinates and more standard ones. We remind the reader
that three dimensional AdS space can be seen as the universal covering of a hyperboloid
$\cal{H}$ of radius $2a$~: 
\be {\cal H}\equiv -U^2-T^2+Y^2+Z^2=-4 a^2 \qquad , \label{HAdS}
\ee
 embedded in the four dimensional (ultrahyperbolic) flat space of metric~: 
\be dS^2=-dU^2-dT^2+dY^2+dZ^2\qquad .
\ee We shall now establish the parametrisation of this quadric leading to the expression
(\ref{dsAdS}) of the metric, and then generalize it to the whole one parameter family of
metrics (\ref{dsm}). To this end, we shall build an auxiliary normal geodesic coordinate
system and relate it both to the
$\{t,x,y\}$ coordinates and to the flat four dimensional coordinates $\{U,T,Y,Z\}$, denoted
generically $\{X^{\alpha}\}$. This will give us the required parametrisation.

The geodesics of AdS space, seen as the hyperboloid $\cal {H}$, are given by the ``great
circles"  defined by the intersection of $\cal {H}$ with two  planes passing through its
center. For our purpose, it is sufficient to restrict ourselves to timelike geodesics.
Indeed, the coordinate transformation is analytic and can thus
 be analytically continued to all types of geodesics,
 once obtained for timelike ones.
 The parametric equations of the  circle passing through the origin [the point of flat
coordinates $(X^{\alpha}(0))=(2a,0,0,0)$] and that of coordinates $(X^{\alpha}_{\star})$ is 
\be { X}^{\alpha}[\lambda ]=X^{\alpha}[0] \cos (\frac{\lambda}{ 2 a})+ {\dot X}^{\alpha}[0]
\sin (\frac{\lambda} { 2 a}) \label{circ}\qquad .
\ee The parameter $\lambda$ is the length measured along this geodesic from the origin to the
point of coordinates $({ X}^{\alpha}[\lambda ])$ and
$({\dot X}^{\alpha}[0])$ are the components of the tangent vector of the geodesic circle at
the origin, given by~:  \be
 {\dot U}[0]=0\ ,\  {\dot T}[0]=\frac{T_{\star}}{\sqrt{4a^2-U_{\star}^2}}\ ,\  
 {\dot Y}[0]=\frac{Y_{\star}}{\sqrt{4a^2-U_{\star}^2}}\ ,\  
 {\dot Z}[0]=\frac{Z_{\star}}{\sqrt{4a^2-U_{\star}^2}}\ .\label{Xdot}
\ee The condition of reality $U_{\star}^2<4a^2$ ensures that the point of coordinates
$(X^{\alpha}_{\star})$ is inside the null cone whose vertex is at the origin $(2a,0,0,0)$.
The distance $s$ between this point and that of coordinates $(X^{\alpha}_{\star})$ is given
by~:
\be
\cos(\frac{s}{2\;a})=2\;a\;U_{\star}\quad .
\ee
 In terms of $\{t,x,y\}$ coordinates, the same geodesic curves read~:
\ba
\lambda+s_0&=&a\arcsin\left(\frac{e^{(x[\lambda]/a)}-C_1\;
C_2}{C_1\sqrt{C_2^2-1}}\right)\quad,\label{geodx1}\\
t[\lambda]+t_0&=&a\arcsin\left(\frac{C_2-C_1e^{-(x[\lambda]/a)}}{\sqrt{C_2^2-1}}\right)
\quad,\label{geodx2}\\
y[\lambda]+y_0&=&-\frac{a}{C_1}\sqrt{C_2^2-1}\cos\left(\frac{t[\lambda]+t_0}a\right)
\quad,\label{geodx3}
\ea where the constants of motion $C_1$ and $C_2$ are the first integrals obtained from the
Killing vector fields
$\xi_1$ and $\xi_2$. The other integration constants $s_0,t_0$ and $y_0$ are chosen such that
the  geodesics start $(\lambda=0)$ at the point  $(t=0,x=0,y=0)$. Indeed, the transitive
isometry group of AdS space allows to make such a choice without loss of generality and even
fix the orientation of the axis such that~: 
\be
 {\dot T}[0]=C_2 \quad ,\quad 
 {\dot Z}[0]=C_2-C_1\quad,\quad   
 {\dot Y}[0]=\pm\sqrt{(C_1^2-1)-(C_2-C_1)^2}\quad. \label{XdotC}
\ee

From equations (\ref{Xdot},\ref{geodx1}--\ref{XdotC}) we obtain, after some algebra, the link
between the flat and  $\{t,x,y\}$ coordinate systems~: 
\ba
\exp(x/a)&=&\frac{(U+Y)^2+(T-Z)^2}{4a^2} \quad ,\nonumber \\
y&=&2\;a\frac{YT+UZ}{(U+Y)^2+(T-Z)^2}\nonumber \quad ,\\
\sin(\frac{t+t_0}a)&=&2\frac{T (Y^2+YU+ Z^2-Z T)+2\; a^2 Z}{[(U+Y)^2+(T-Z)^2]\sqrt{Y^2+Z^2}}
\quad . \label{trans}
\ea The initial values $t_0$ and $y_0$ are undefined in terms of the  $X^{\alpha}$
coordinates, but if we use the standard parametrisation of AdS space defined by~: 
\ba
 U=2a\cosh(r)\cos(\tau) \quad &,&\quad T=2a\cosh(r)\sin(\tau) \quad,\quad \nonumber \\
Y=2a\sinh(r)\cos(\theta) \quad &,&\quad Z=2a\sinh(r)\sin(\theta) \quad,\label{emAdS}
\ea we obtain 
\be y_0=-a\frac{Y_{\star}}{T_{\star}-Z_{\star}}\quad,\quad t_0=a\;\theta_{\star} \quad ,
\ee and the coordinate transformation (\ref{trans}) can be rewritten~:
\ba
\exp(x/a)&=&\cosh(2r)+\cos(\theta + \tau)\sinh(2r)\quad ,\label{transAdS1} \\
y\exp(x/a)&=&a \sinh(2r)\sin(\theta+\tau)\quad ,\label{transAdS2} \\
\sin(\frac ta+\theta)&=&\frac{\cosh(2r)\cos(\theta
+\tau)\sin(\tau)+\cos(\tau)\sin(\theta+\tau)+\sinh(2r)\sin(\tau)}
{\cosh(2r)+\cos(\theta+\tau)\sinh(2r)}\quad .\label{transAdS3}
\ea The last equation may be expressed in  the more simple,  but equivalent, form~:
\be
\tan [\frac12(\frac ta + \theta-\tau)]=\exp(-2r) \tan [\frac12 (\theta+\tau)]\quad .
\label{transAdS4} \ee

In eqs (\ref{geodx1}--\ref{geodx3}) the natural range of the variables $\lambda$ and $t$ is
the interval $[0,2a\pi]$, whereas in eq. (\ref{circ}) the range of $\lambda$ is obviously
$[0,4a\pi]$. This reflects the fact that the metric (\ref{dsAdS}), with $t$ restricted to
$[0,2a\pi]$  actually describes a projective AdS space. This is also clear from eqs.
(\ref{transAdS1}--\ref{transAdS3}) which show that the coordinates $\{t,x,y\}$ are left
unchanged with respect to the transformation $(\theta,\tau)\mapsto (\theta + \pi,\tau +
\pi)$, i.e. $X^{\alpha}_\star \mapsto -X^{\alpha}_\star$. However, if  we let $t$ run over the
whole real line, we obtain the usual (universal covering) AdS space.\\  In terms of the
$\{\tau, \theta,r\}$ coordinates, the AdS metric takes the well known form~: \be d
\sigma^2 = 4 a^2[- \cosh^2(r) d\tau^2 +dr^2 + \sinh^2(r) d\theta^2] \quad  , 
\ee adapted to the Killing vectors~:
\be
\partial_{\tau}=a(\frac 12 \xi_1 - \xi_3 +\xi_4)\quad,\quad
\partial_{\theta}=a(\frac 12 \xi_1 - \xi_3 -\xi_4)\quad .\label{kilAdS} \ee This suggests to
perform a similar transformation for arbitrary values of the $\mu$ parameter, but the
resulting metric presents a conical singularity. For arbitrary values of $\mu$, the
coordinate transformation (\ref{transAdS1}-\ref{transAdS4})  generalizes as follows
\cite{RTi}~:  
\ba
\exp(x/a)&=&\cosh(2r)+\cos(\frac{\theta + \tau}{\mu})\sinh(2r)\quad ,\label{ctrans1}\\
y\exp(x/a)&=& \mu a \sinh(2r)\sin(\frac{\theta + \tau}{\mu})\quad ,\label{ctrans2}\\
\tan [\frac12(\frac  t{\mu a} + \frac{\theta - \tau}{\mu})]&=&\exp(-2r) \tan [\frac 12
(\frac{\theta + \tau}{\mu})]\quad , \label{ctrans3}
\ea
 and leads to the metrics~: 
\be
 d \sigma _{\mu}^2= 4 a^2[- \cosh^2(r) d\tau^2 +dr^2 + \sinh^2(r) d\theta^2
+\frac{1-\mu^2}{\mu^2} \sinh^2(r) \cosh^2(r) (d\tau +d\theta)^2] \quad  . 
\ee The $\tau$ and $\theta$  coordinates are now adapted to the Killing vectors~:
\be a(\frac 12 \xi_1 -\frac1{\mu^2} \xi_3 +\xi_4)=\partial_{\tau}\quad,\quad a(\frac 12 \xi_1
-
\frac1{\mu^2}\xi_3 -\xi_4)=\partial_{\theta}\quad .\label{kil}
\ee Introducing  the new angular variable $\phi=(\theta+\tau)/\mu$, we finally obtain~:   
\be d\sigma _{\mu}^2= 4 a^2\left[-d\tau^2 +dr^2 +[\sinh^2(r) +(1-\mu ^2) \sinh^4(r)]
d\phi^2-2
\mu \sinh^2(r) d\tau d\phi\right] \quad ,\label{dsfin}
\ee which, for $\mu^2=2$, yields the coordinate transformation and  metric  obtained fifty
years ago by  G\"odel \cite{God1}. Note that   the angle $\phi$ has to vary between $0$ and
$2\pi$, otherwise the coordinate transformation (\ref{ctrans1}-\ref{ctrans3}) is no longer
one-to-one, and the metrics  present  conical singularities. This explains the presence of 
the parameter $\mu$ in the argument of the trigonometric functions used in eqs
(\ref{ctrans1}-\ref{ctrans3}) and as prefactor of the Killing vector $\xi_3$ in eqs
(\ref{kil}). As a consequence, in terms of $\{\tau,r,\theta\}$ coordinates, we have to
identify points whose $\theta$ values  differ by $2\pi\mu$.

 The G\"odel metric is known to contain closed timelike curves, whereas AdS space (at least
its universal covering) not.  Examination of eq. (\ref{dsfin}) shows that the closed
$\phi$ curves become timelike for values of $r> r_c(\mu)=\frac 12 \log \frac {\mu +1 }
{\mu -1}$, where $\mu$ is assumed to be positive. Moreover, as the isometry group acts
transitively on the space--time, such closed curves pass through all points. Hence,
there exist closed timelike curves for all values of
$\mu>1$.  This can be visualized geometrically by reinterpreting the metrics (\ref{dsm}) as a
family of tensor fields
$\mbox{\bf g}_{\mu}$ on an AdS background space and comparing the different apertures of the
lightcones ${\cal C}_{\mu}\equiv\mbox{\bf g}_{\mu}(\vec v,\vec v)=0$ as  function of $\mu$
and their location. With respect to the basis
$\{\vec e_{0},\vec e_{1},\vec e_{2}\}$, dual of the triad (\ref{triad}), these cones have
everywhere the same shape. They are stretched in the $\vec e_{2}$ direction for
$\mu>1$, and squashed for $\mu<1$ (see Fig.1). On the other hand, if  this frame is boosted
to the basis $\{\vec e_{\tau}=\mbox{\rm sech}(r)\partial_{\tau},
\vec e_{r}=\partial_{r},\vec e_{\theta}=\mbox{\rm csch}(r)\partial_{\theta}\}$, the
cones appear deformed with respect to the (invariant) AdS lightcone (Figs. 1-3). As the
$r$ coordinate increases, the cones with $\mu<1$ narrow down and tilt towards the $(-\vec
e_{\theta})$ direction; in the limit of $r \mapsto \infty $ they coincide with the AdS
null direction $\vec e_{\tau}+\vec e_{\theta}$. The cones with $\mu>1$, as for them, open out
in the  $\vec e_{\theta}$ direction; once $r>r_c(\mu)$, they include the $\vec
e_{\theta}$ axis. As a consequence, the circles of radius $r>r_c(\mu)$ have their
tangent vectors everywhere inside  ${\cal C}_{\mu}$, thus yielding closed timelike curves.

From eq. (\ref{dsfin}) it is easy to compute the geodesics and extend the analysis  already
performed  for AdS  and G\"odel's metrics \cite{geod,HwEl}. We find that for $\mu>1$, the
qualitative behavior of the geodesics are the same as in G\"odel's space. The light geodesics
starting from a point spiral up, diverging up to a  caustic circle of radius $r=\frac 12
\log \frac {\mu +1 } {\mu -1}$ (measured on a surface of constant $\tau$)  and then
reconverge. For $\mu=1$, the closed timelike curves are pushed to infinity, but there still
exist Cauchy horizons, a well known property of AdS space. For $\mu<1$ we have not detected
any causal pathology.  

Another way to understand the presence of closed time lines is to consider isometric
embeddings of the metrics (\ref{dsfin}) in flat spaces.  Though the three dimensional AdS
space can easily be embedded in four dimensional flat space [eq.(\ref{emAdS})], the best
result we have obtained for the other spaces of the family is the almost obvious, but global,
embedding in seven dimensional flat spaces of metrics~: 
\be dS^2=-dU^2-dT^2+dY^2+dZ^2+\varepsilon (dA^2+dB^2-dC^2)\qquad ,\quad\varepsilon=\pm 1=
\mbox{\rm sign}(1-\mu^2) \label{DS}
\ee via the parametrisation~: 
\ba
 U=2\;\mu\;a\cosh(r)\cos(\frac{\tau}{\mu}) \quad &,&\quad
T=2\;\mu\;a\cosh(r)\sin(\frac{\tau}{\mu})
\quad,\nonumber \\  Y=2\;\mu\;a\sinh(r)\cos(\frac{\theta}{\mu}) \quad &,&\quad
Z=2\;\mu\;a\sinh(r)\sin(\frac{\theta}{\mu})\nonumber
\\ A=a\;\sqrt{|1-\mu^2|}  \sinh(2 r) \cos(\frac{\tau + \theta}{\mu})\quad &,&\quad 
B=a\;\sqrt{|1-\mu^2|} \sinh(2 r) \sin(\frac{\tau + \theta}{\mu})\quad,\nonumber \\
C=a\;\sqrt{|1-\mu^2|}  \cosh(2 r) \quad &.&
\ea This embedding can be seen as the intersection of the four quadratic surfaces~: 
\ba {\cal H}_1&\equiv&-U^2-T^2+Y^2+Z^2=-4\;a^2 \quad, \nonumber\\ {\cal
H}_2&\equiv&C^2-A^2-B^2=a^2\;|1-\mu^2| \quad , \nonumber\\ {\cal P}_1&\equiv& A - \frac
{\sqrt{|1-\mu^2|}}{2\;a\;\mu ^2} (U\;Y-T\;Z ) =0 \quad ,
\nonumber \\
 {\cal P}_2&\equiv& B - \frac {\sqrt{|1-\mu^2|}}{2\;a\;\mu ^2} (U\;Z + T\;Y) =0
\quad. 
\ea We easily see that, when $\mu ^2=1$, we recover the standard embedding of AdS space as the
hyperboloid ${\cal H}$ (\ref{HAdS}), here in the four dimensional subspace $A=B=C=0$. In the
other cases, we see that both cylindrical surfaces ${\cal H}_1$ and ${\cal H}_2$ possess non
contractible closed curves. The first is topologically equivalent to the cartesian product of
the AdS hyperboloid ${\cal H}$ with $ \R  ^3$  and can be unwinded in the
$\tau$-direction, thereby removing the closed time like curves. The second, ${\cal H}_2$, 
cannot be unwinded without introducing a singularity in the three dimensional manifolds, which
do not remain homeomorphic to $\R  ^3$ at $r =0$. The circles  generated by varying the
$\phi$ coordinate  are  always spacelike when
$\varepsilon=+1$, that is, for $\mu^2<1$; for $\varepsilon=-1$, $\mu^2>1$, they become
timelike  when, in eq. (\ref{DS}), the term 
$dA^2+dB^2$ dominates $dY^2+dZ^2$, i.e. once their radii are large enough.

The main interest of the previous analysis resides in the geometric picture  we have
obtained of the G\"odel--like universes. On the one hand, we have shown how these geometries
can be interpreted as resulting from the squashing of AdS lightcones. On the other hand, we
have obtained a simple global embedding of these spaces, showing clearly the
unavoidability of  closed timelike curves for spaces with $\mu^2>1$, AdS space ($\mu^2=1$)
appearing as the limiting case. \\
\strut\\
{\bf Acknowledgments} We are grateful to J. Demaret for helpful references, and M. Henneaux
and M. Lubo for interesting comments and discussions. We also thank the referee whose kind
suggestions have drawn our attention to \cite{RTi}, in which the coordinate
transformation (\ref{ctrans1}--\ref{ctrans3}) also appears.

\vfill
\newpage
\section*{Figure caption}
\begin{description}
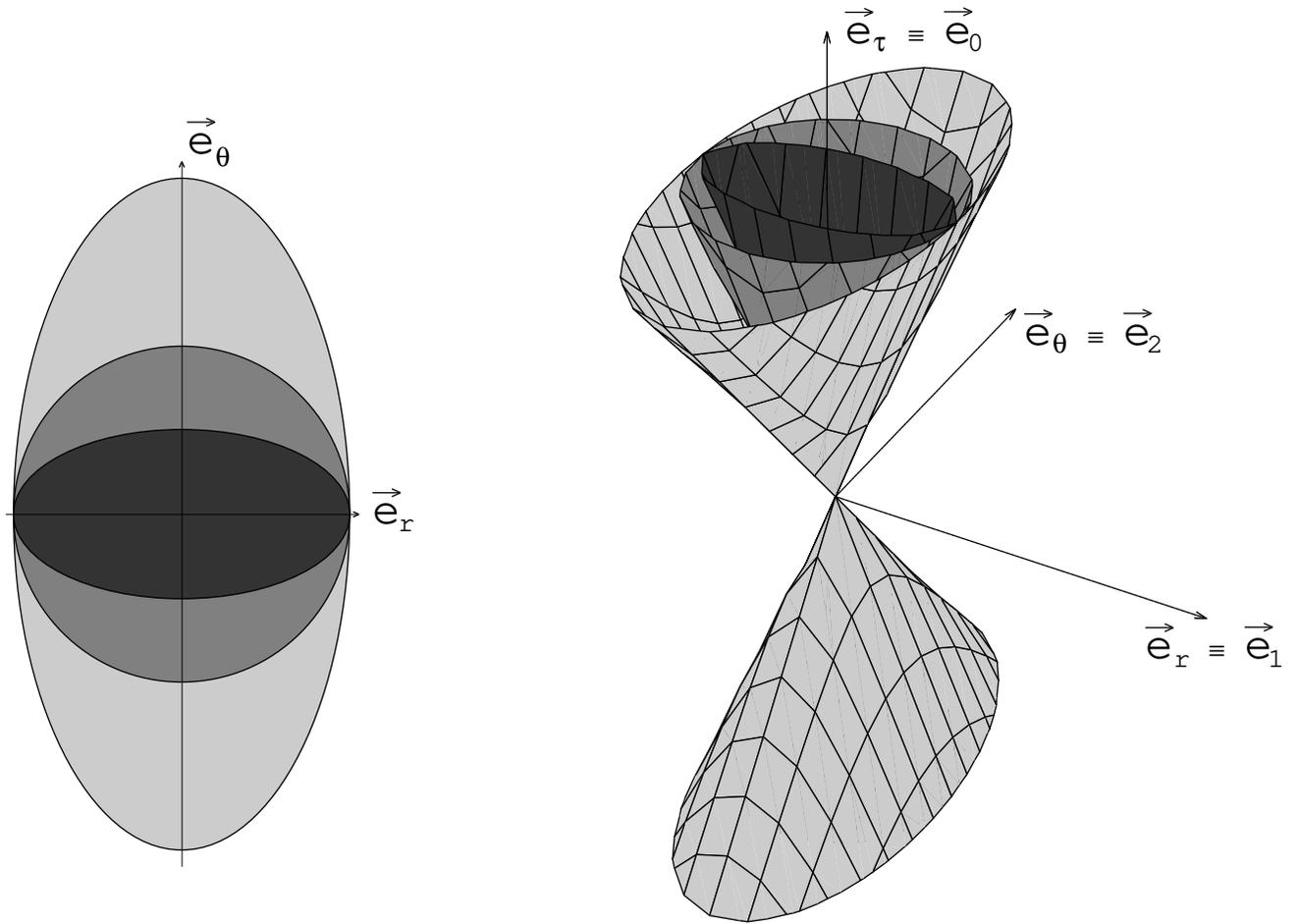
\item[Figure 1.] Right hand side~: The cone ${\cal C}_{1/2}$ (in dark gray) and the cone
${\cal C}_{2}$ (in pale gray), respectively inside and around the AdS lightcone ${\cal
C}_{1}$ (in medium gray), at
$r=0$, where the frames  $\{\vec e_{0},\vec e_{1},\vec e_{2}\}$ and $\{\vec e_{\tau},\vec
e_{r},\vec e_{\theta}\}$ are identical. Left hand side~: a plane section orthogonal to
$\vec e_{\tau}$ of the three nested cones. 

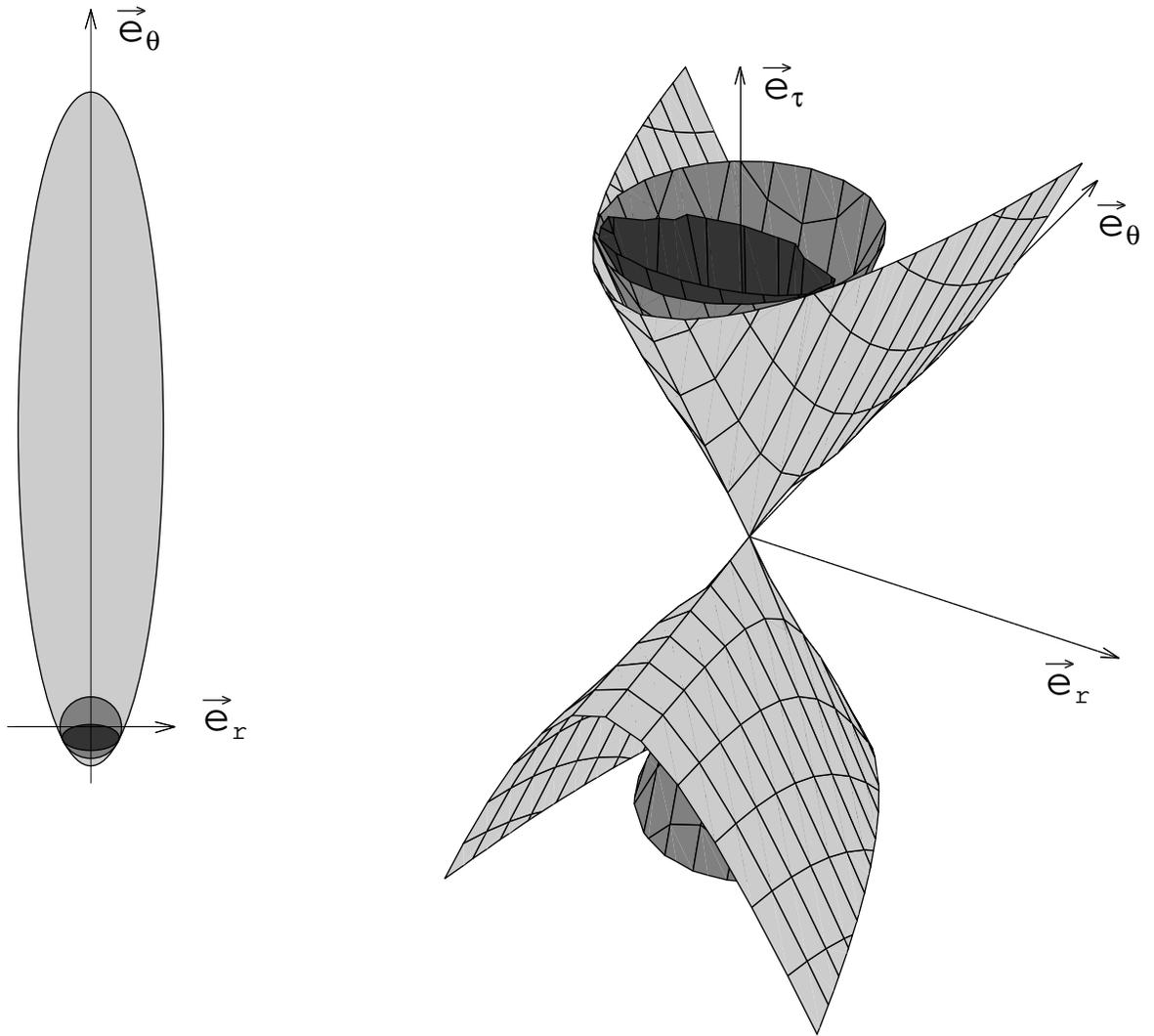
\item[Figure 2.]  Right hand side~: The cone ${\cal C}_{1/2}$ (in dark gray) and the cone
${\cal C}_{2}$ (in pale gray), respectively inside and around the AdS lightcone ${\cal
C}_{1}$ (in medium gray), at
$r=0.5<r_c(2)\approx 0.549$. Left hand side~: a plane section orthogonal to $\vec
e_{\tau}$ of these three nested cones.

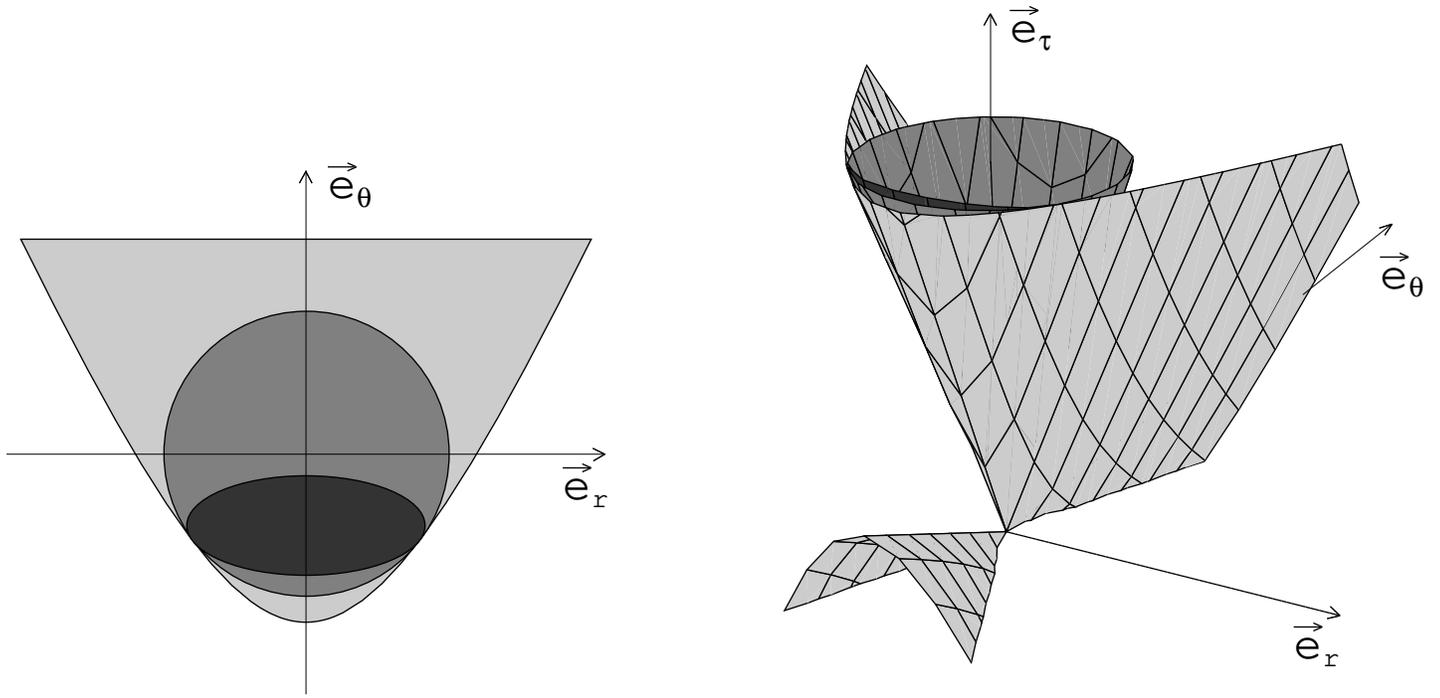
\item[Figure 3.]  Right hand side~: The cone ${\cal C}_{1/2}$ (in dark gray) and the cone
${\cal C}_{2}$ (in pale gray), respectively inside and around the AdS lightcone ${\cal
C}_{1}$ (in medium gray), at
$r=0.7>r_c(2)\approx 0.549$. Here  only the parts above the plane $[\vec e_{r},\vec
e_{\theta}]$ have been drawn, showing that the  $\vec e_{\theta}$ direction is now  inside
${\cal C}_{2}$.
 Left hand side~: a plane section orthogonal to $\vec e_{\tau}$ of these three nested cones,
on which the branch of hyperbola representing the intersection of the plane with the lower
half cone ${\cal C}_{2}$ has been suppressed.
\end{description}

\begin{thebibliography}{999}

\bibitem{God1} K. G\"odel, Rev. Mod. Phys., {\bf 21} (1949) 447; in {\sl Proceedings of the
international Congress of Mathematicians, Cambridge, Massachusetts, U.S.A.,(1950)}, ed. 
Kraus Reprint Limited, Nenden/Liechtenstein (1967), page 175.

\bibitem{God2}  K. G\"odel, in {\sl Albert Einstein~:  Philosopher--Scientist}, ed. P.A.
Schilpp, The Library of Living Philosophers, Inc., Evanston, Illinois (1949) page 557\\
 A. Einstein, in {\sl Albert Einstein~:  Philosopher--Scientist}, ed. P.A. Schilpp, The
Library of Living Philosophers, Inc., Evanston, Illinois (1949) page 665\\ H. Stein,
Philosophy of Science, {\bf 37} (1970) 589.

\bibitem{KSHM} D. Kramer, H. Stefani, E. Herlt, M. MacCallum, {\sl Exact Solutions of
Einstein's Field Equations},Cambridge University Press, Cambridge (1980).

\bibitem{HwEl} S. W. Hawking, G.F.R. Ellis, {\sl The Large Scale Structure of Space-time},
Cambridge University Press, Cambridge (1973)  chapter 5.

\bibitem{Hw} S. W. Hawking, Phys. Rev. {\bf D44} (1991) 3802.

\bibitem{Na} P. Nahin, {\sl Time Machines}, Spinger--Verlag (1993).

\bibitem{BSMcC} W. B. Bonnor, N. O. Santos, M. A. H. MacCallum, Class. Quantum Grav. {\bf 15}
(1998) 357.

\bibitem{LRS} M. Lubo, M. Rooman, Ph. Spindel, {\sl 2+1 Dimensional stellar models} (in
preparation).

\bibitem{RTh} A. K. Raychaudhuri, S. N. Guha Thakurta, Phys. Rev. {\bf D22} (1980) 802.

\bibitem{RTi} M. J. Rebou\c cas, J. Tiomno, Phys. Rev. {\bf D28} (1983) 1251.

\bibitem{BD} J. D. Barrow, M. P. D\c abrowski, {\sl G\"odel Universes in String Theory}, 
gr-qc 9803048.

\bibitem{geod} S. Chandrasekhar, J. P. Wright, Proc. Nat. Acad. Sci., {\bf 48} (1961) 341.


\end{thebibliography}
\end{document}